\documentclass[9pt,twocolumn,twoside]{osajnl}
\journal{ol} % Choose journal (ao,jocn,josaa,josab,ol,optica,pr)

\setboolean{shortarticle}{true}
\usepackage{gensymb}
\usepackage{siunitx}
\usepackage{physics}

\title{Spectrally separable photon-pair generation in dispersion engineered thin-film lithium niobate}

\author[1, 5, \textdagger]{C.~J.~Xin}
\author[2, 6, \textdagger]{Jatadhari~Mishra}
\author[3]{Changchen~Chen}
\author[1]{Di~Zhu}
\author[1]{Amirhassan~Shams-Ansari}
\author[2]{Carsten~Langrock}
\author[1, 4]{Neil~Sinclair}
\author[3]{Franco~N.~C.~Wong}
\author[2]{M.~M.~Fejer}
\author[1]{Marko~Lončar}

\affil[1]{John~A.~Paulson School of Engineering and Applied Sciences, Harvard University, Cambridge, Massachusetts~02138, USA}
\affil[2]{Edward~L.~Ginzton Laboratory, Stanford University, Stanford, California~94305, USA}
\affil[3]{Research Laboratory of Electronics, Massachusetts Institute of Technology, Cambridge Massachusetts~02139, USA}
\affil[4]{Division of Physics, Mathematics and Astronomy, and Alliance for Quantum Technologies, California Institute of Technology, Pasadena, California~91125, USA}
\affil[5]{email: cxin@g.harvard.edu}
\affil[6]{email: jmishra@stanford.edu}

\begin{abstract}
    Existing nonlinear-optic implementations of pure, unfiltered heralded single-photon sources do not offer the scalability required for densely integrated quantum networks. 
    Additionally, lithium niobate has hitherto been unsuitable for such use due to its material dispersion.
    We engineer the dispersion and the quasi-phasematching conditions of a waveguide in the rapidly emerging thin-film lithium niobate platform to generate spectrally separable photon pairs in the telecommunications band. 
    Such photon pairs can be used as spectrally pure heralded single-photon sources in quantum networks. 
    We estimate a heralded-state spectral purity of~$\mathbf{{>}94\%}$ based on joint spectral intensity measurements.
    Further, a joint spectral phase-sensitive measurement of the unheralded time-integrated second-order correlation function yields a heralded-state purity of $\mathbf{\qty(86 \pm 5) \%}$. 
\end{abstract}

\setboolean{displaycopyright}{true}

\begin{document}
\maketitle

\noindent 
Thin-film lithium niobate (TFLN) can host many active devices of interest for scalable room-temperature and high-bandwidth photonic quantum technology, including state-of-the-art electro-optic modulators~\cite{Xu:22}, optical frequency converters~\cite{Wang:18}, photon detectors~\cite{Desiatov:19}, and optical $\chi^{(2)}$-based photon-pair sources~\cite{Elkus:19,Ma:2020,Javid:21}.
Properly engineered, a photon-pair source can also be used as a single-photon source by heralding the existence of one photon through detection of the other one~\cite{Eisaman:11, Mosley:08}.
However, LN-based sources typically yield photon pairs in spectrally entangled biphoton states due to material dispersion.
Such entanglement  results in incoherence between different frequency components upon detection of the heralding photon, and causes the heralded photon to be produced in a mixed state~\cite{URen:05}.
This renders the heralded photon unsuitable for quantum interference, and thus unusable for optical quantum computing and networking.

Existing solutions to this problem include strong spectral filtering of the heralded photon~\cite{MeyerScott:17}, nonlinear interaction in cavities~\cite{Luo:15, Guo2016}, or using a different nonlinear medium with a more suitable material dispersion~\cite{Mosley:08, Graffitti2018, Chen:19}.
However, such techniques come with the problems of reduced photon generation rates, compromised heralding efficiencies, or lack of thin-film wafer availability for scalable, integrated nanophotonics. 

In this work, we demonstrate joint spectral separability in TFLN by exploiting the dispersion control enabled by sub-wavelength photon confinement in thin films.
By optimizing the TFLN waveguide geometry, we overcome the limitations imposed by LN's material dispersion~\cite{Jankowski2021}.
Additionally, we spatially apodize the quasi-phasematching (QPM) grating to suppress side lobes present in the phasematching functions (PMF) of uniform QPM gratings, and thereby producing a Gaussian PMF~\cite{Huang:06}.
Thus, we experimentally realize a monolithically integrable photon-pair source on TFLN, with heralded-state spectral purity $P_\text{spectral} > 94\%$ inferred from the measured joint spectral intensity ~\cite{Chen:17}. 
To further characterize the heralded-state purity, we measure the unheralded time-integrated second-order correlation function $g^{(2)} = 1.86 \pm 0.05$ of the signal photons.
This provides an estimate of the heralded-state purity $P$, where $P \simeq g^{(2)} - 1$, and $P = 1$ corresponds to complete spectral separability, including spectral phase~\cite{Christ:11}.

Photon-pair generation in LN takes place via spontaneous parametric downconversion (SPDC), where a pump photon of center frequency $\omega_{p}$ is downconverted to a signal and an idler photon of center frequencies $\omega_{s}$ and $\omega_{i}$, respectively, such that energy is conserved, i.e. $\omega_p = \omega_s + \omega_i$.
This process is described by the semi-classical Hamiltonian 
\begin{equation}
    \hat{H} = \iint \dd{\omega_i} \, \dd{\omega_s} \, f\qty(\omega_i, \omega_s) \, \hat{a}_i^\dagger \hat{a}_s^\dagger + \text{h.c.} 
    \label{eq1},
\end{equation}
where $f\qty(\omega_i, \omega_s) = \alpha_p(\omega_{i} + \omega_{s}) \, \phi(\omega_{i}, \omega_{s})$ is the joint spectral amplitude (JSA) of the biphoton state;
$\alpha_p$ is the pump envelope function (PEF), and $\phi$ is the PMF.
The joint spectral intensity (JSI) is 
$\abs{f\qty(\omega_i, \omega_s)}^2 = \abs{\alpha _p(\omega _{i} + \omega _{s})}^{2} \abs{\phi(\omega_{i}, \omega _{s})}^{2}$.
A spectrally separable biphoton state, which is necessary for producing a spectrally pure heralded state, requires $f(\omega_{i}, \omega_{s}) = f_{i}(\omega_{i}) f_{s}(\omega_{s})$.
In this work, we design for spectral separability up to second order in phase mismatch by satisfying the group-velocity mismatch (GVM) condition $\qty(v^{-1}_{g,s} - v^{-1}_{g,p})/\qty(v^{-1}_{g,p} - v^{-1}_{g,i}) \geq 0$, where $v_{g,m}$ ($m = s, i, p$) are the group velocities. 
This condition is satisfied if the pump group velocity lies between that of the signal and idler, or is exactly equal to either of those. While this ensures the separability of the JSI, the spectral phase correlations in a JSA resulting from the product of a Gaussian pump and a Gaussian PMF can be eliminated by standard pulse shaping techniques that satisfy $2\beta_{t} + \beta_{p}/2 = 0$, where $\beta_{t}$ and $\beta_{p}$ represent the group-delay dispersion the pump experiences before and in the waveguide, respectively~\cite{URen:05}. 

% Design figure
\begin{figure}[!htb]
    \centering
    \includegraphics[width=0.9\linewidth]{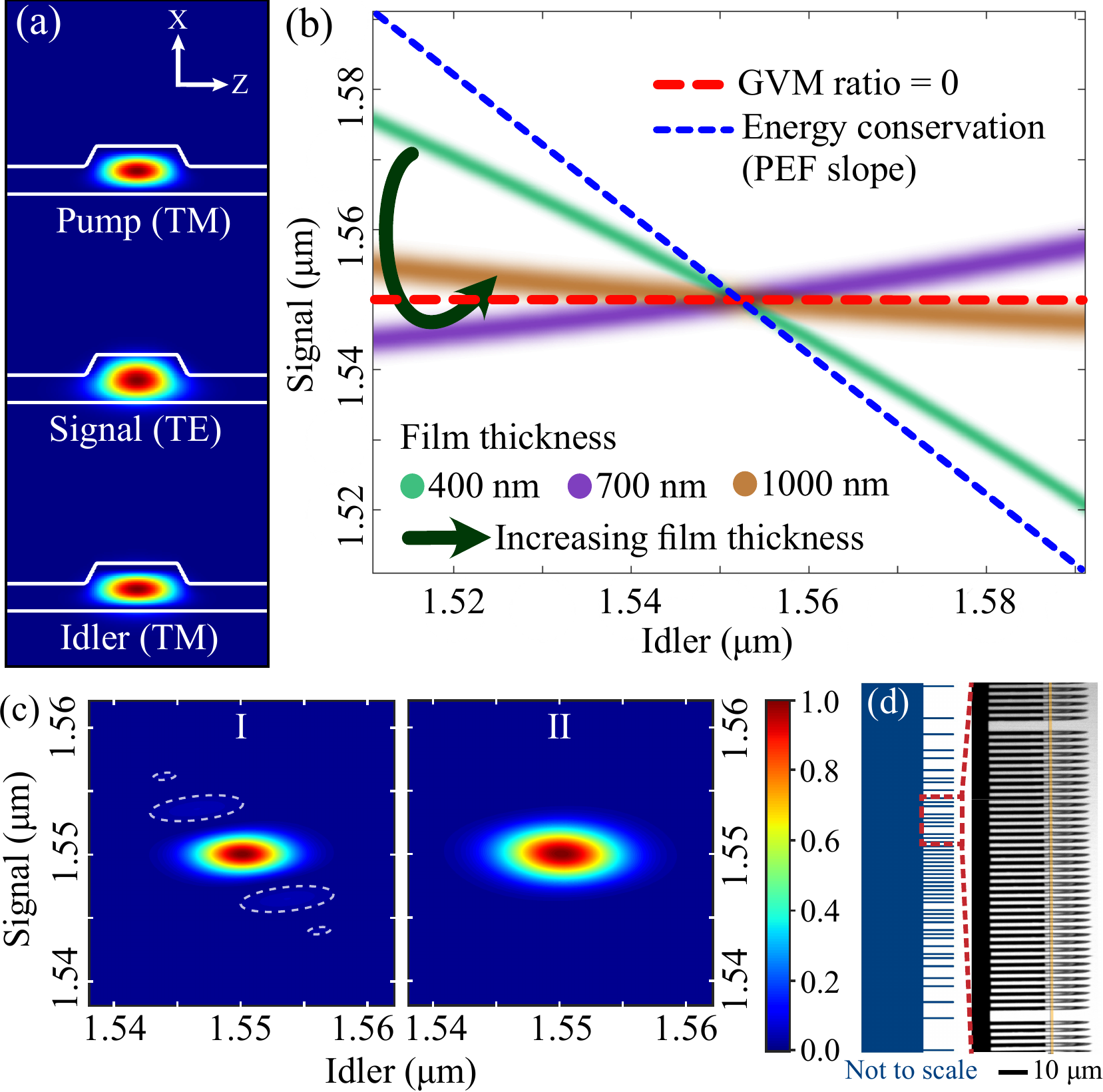}
    \caption{
        Device design.
        (a)~Normalized cross-sectional mode intensities of the pump at 775 nm wavelength and both signal and idler at 1550 nm wavelength. TFLN is outlined in white with air on top and silica cladding underneath, with nominal waveguide dimensions of top width 1200~nm, film thickness 700~nm, etch depth 300~nm, sidewall angle 62$\degree$.
        (b)~Simulated PMF for varied film thickness, other dimensions are same as above; waveguide length 5 mm.
        The lines representing zero GVM ratio and slope of the PEF are represented by red and blue dashed lines, respectively.
        (c)~Simulated normalized JSIs for waveguide geometry in (a) given by the overlap of Gaussian pumps of FWHMs 0.8~THz and 1.0~THz, and simulated PMFs for (I) non-apodized QPM grating (resulting in a sinc PMF; dashed outline around side lobes for emphasis) and (II) Gaussian-apodized QPM grating (resulting in a Gaussian PMF), respectively, thus optimizing the respective $P_\text{spectral}$ to $90.5\%$ and $99.8\%$.
        (d)~Deleted-domain Gaussian-apodized poling electrode pattern (left). 
        Two-photon microscope image of the poled film (right); overlaid yellow line represents the waveguide while black and grey regions show the electrodes and inverted domains, respectively.
    }
    \vspace{-2ex}
    \label{fig:design}
\end{figure}
Our design uses waveguide geometry-induced dispersion to satisfy the GVM condition for generating a spectrally separable biphoton state at telecommunications wavelengths.
Due to its higher fabrication tolerance, a type-II phasematching scheme is used~\cite{Jankowski2021}, where the pump beam is in a quasi-TM$_{00}$ mode, and the signal and idler beams are in quasi-TE$_{00}$ and quasi-TM$_{00}$ modes, respectively (Fig.~\ref{fig:design}a).
Since the PEF slope is fixed by energy conservation to be negative, the PMF slope must be zero or positive to generate a spectrally separable biphoton state. 
Appropriate dispersion is achieved by tuning the TFLN waveguide geometry, with the film thickness being the most sensitive parameter~(Fig.~\ref{fig:design}b).
Of the investigated film thicknesses, the intermediate value of 700~nm provides the desired PMF slope.

For maximum separability, a Gaussian PMF is generated using deleted-domain Gaussian apodization of the periodic poling, as opposed to sinc-squared profile obtained from a uniform (i.e. non-apodized) grating. 
Detailed explanations of the implementation of the deleted-domain apodization scheme can be found in Refs.~\cite{Huang:06, Jankowski2021}.
The simulated JSIs for non-apodized and apodized PMFs are shown in Fig.~\ref{fig:design}c panels I and II, respectively. 
The apodized poling electrode layout is schematized in Fig.~\ref{fig:design}d (left), while part of the poled film prior to waveguide fabrication is shown in the inset (right).
Poled waveguides are fabricated on an $x$-cut 700~nm MgO-doped TFLN on insulator die following Ref.~\cite{Wang:18}.
Air-clad waveguides with Gaussian-apodized as well as non-apodized QPM gratings are fabricated for comparison; phasematching was tuned across the die by poling each device with a slightly different period near $3.25~\si{\micro\m}$.

% SFG measurement figure
\begin{figure}[!htb]
    \centering
    \includegraphics[width=0.85\linewidth]{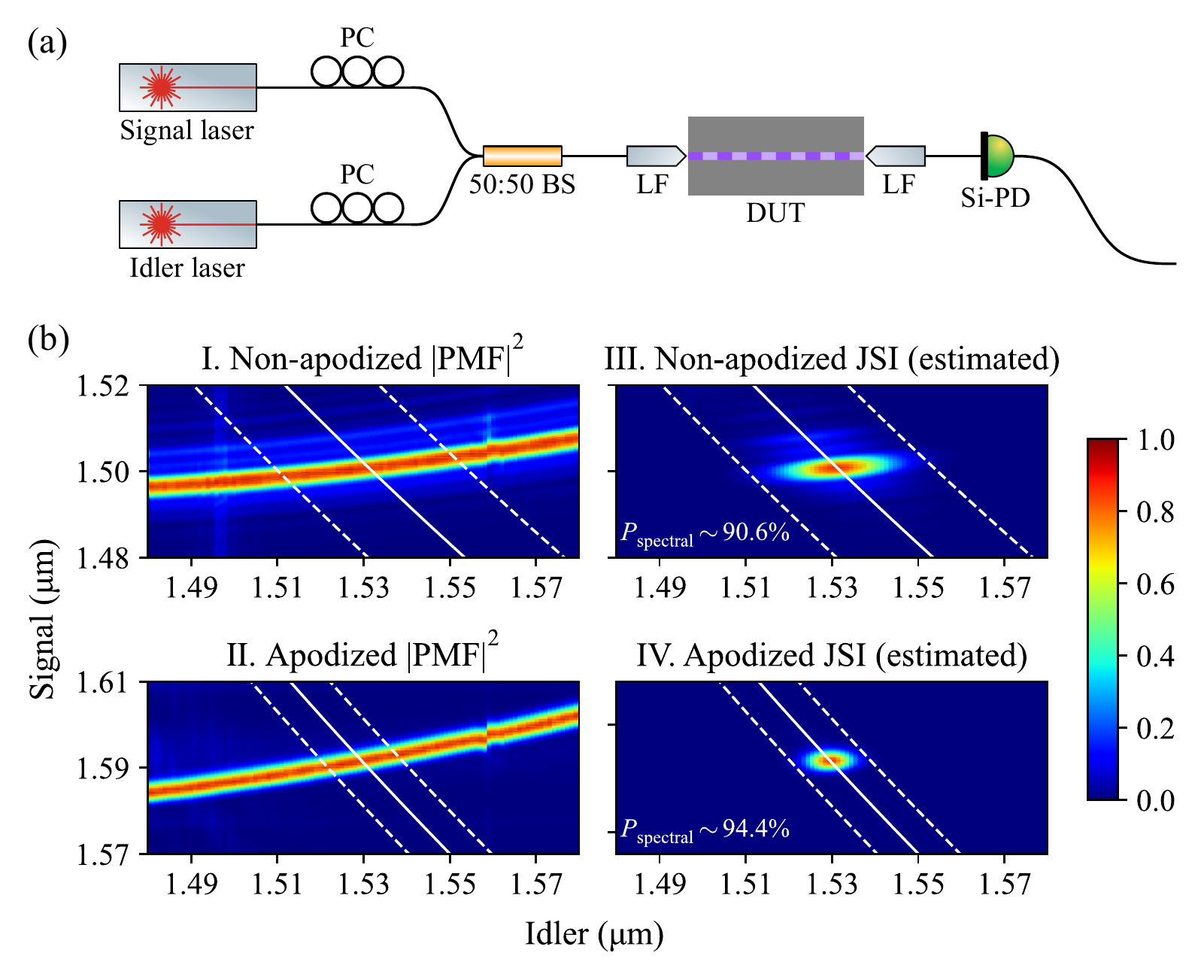}
    \caption{
        PMF characterization via SFG. 
        (a)~Measurement setup. 
        PC,~polarization controller; 
        BS,~beam~splitter; 
        LF,~lensed fiber; 
        DUT,~device under test; 
        Si-PD,~silicon photoreceiver. 
        (b)~Normalized measured PMFs for a waveguide with (I)~non-apodized periodic poling and (II)~Gaussian-apodized periodic poling. 
        Estimated normalized joint spectral intensities with optimal purities of ${\sim}90.6\%$ and ${\sim}94.4\%$ are obtained for the device with (III)~non-apodized periodic poling and the device with (IV)~Gaussian-apodized periodic poling, respectively. 
        Overlaid solid white lines indicate the center wavelengths, $\lambda_p$, of the optimal pump spectra needed to maximize JSI purities, while the dashed lines indicate the FWHM, $\Delta \nu_p$, envelopes of the optimal pump spectra.
        For the device with non-apodized QPM, $\lambda_p = 756~\si{\nm}$ and $\Delta\nu_p = 2.83~\si{\THz}$, whereas for the device with apodized QPM, $\lambda_p = 780~\si{\nm}$ and $\Delta\nu_p = 1.20~\si{\THz}$.
    }
    \vspace{-2.5ex}
    \label{fig:sfg_expt}
\end{figure}

To characterize our devices, we first measure the PMF via sum-frequency generation (SFG). 
The SFG process creates a field at frequency $\omega_p$ by combining signal and idler fields with frequencies $\omega_s$ and $\omega_i$, respectively, and shares the same phasematching conditions as the SPDC process involving these photons.
Assuming that the powers of the idler and signal fields $\mathcal{P}\qty(\omega_i)$ and $\mathcal{P}\qty(\omega_s)$, respectively, are constant and undepleted, the power $\mathcal{P}\qty(\omega_p)$ generated at the sum-frequency, $\omega_p$, obeys $\abs{\phi(\omega_i, \omega_s)}^2 \propto \mathcal{P}(\omega_p)/[\mathcal{P}\qty(\omega_i) \mathcal{P}\qty(\omega_s)]$. 
Accordingly, we extract $\abs{\phi(\omega_i, \omega_s)}^2$ by tuning $\omega_i$ and $\omega_s$ and measuring $\mathcal{P}\qty(\omega_p)$.
Signal and idler beams are generated using two tunable continuous-wave telecommunications-band lasers, combined using a 50:50~beamsplitter, and end-fire coupled into the poled waveguides (Fig.~\ref{fig:sfg_expt}a). 
The generated SFG signal is collected using a lensed fiber and directed to a silicon photoreceiver.  

Our measurement yields PMFs with the desired positive slopes, indicating that the targeted waveguide dispersion is achieved (Fig.~\ref{fig:sfg_expt}b, panels I and II). 
Moreover, side lobes visible in the PMF of the waveguide with non-apodized QPM (Fig.~\ref{fig:sfg_expt}b, panel~I) are suppressed in that with deleted-domain Gaussian apodization (Fig.~\ref{fig:sfg_expt}b, panel~II). 
We note that these two waveguides were chosen based on fabrication quality, but had slightly different poling periods, and thus differ in their phasematching.
For each measured PMF, we estimate the JSI by multiplying the measured PMF with an optimal Gaussian pump envelope that maximizes the heralded-state spectral purity, $P_\text{spectral}$.
We find that the resulting $P_\text{spectral}$ for both exceed $90\%$ (Fig.~\ref{fig:sfg_expt}b, panels III and IV), with it being higher for the device using the Gaussian-apodized QPM than that using the non-apodized QPM, as expected. 
Deviations from simulations in Fig.~\ref{fig:design} are due to small differences between nominal and fabricated waveguide geometries, small asymmetries in apodized QPM, and different phasematched wavelengths.
In bulk LN, such dispersion engineering is not possible, and biphoton states in the telecommunications band with $P_\text{spectral}$ ${>}90\%$ are only achievable upon extreme filtering, which in turn greatly reduces the photon rates~\cite{Laudenbach:16}.
Additionally, we characterized the normalized SFG efficiency of the Gaussian apodized device, given by the ratio $\mathcal{P}(\omega_p)/[\mathcal{P}\qty(\omega_i) \mathcal{P}\qty(\omega_s)]$ at the phasematching peak, to be 0.24~$\text{W}^{-1}$, which is also in good agreement with simulation.

% Setup figure
\begin{figure}[!htb]
    \centering
    \includegraphics[width=0.9\linewidth]{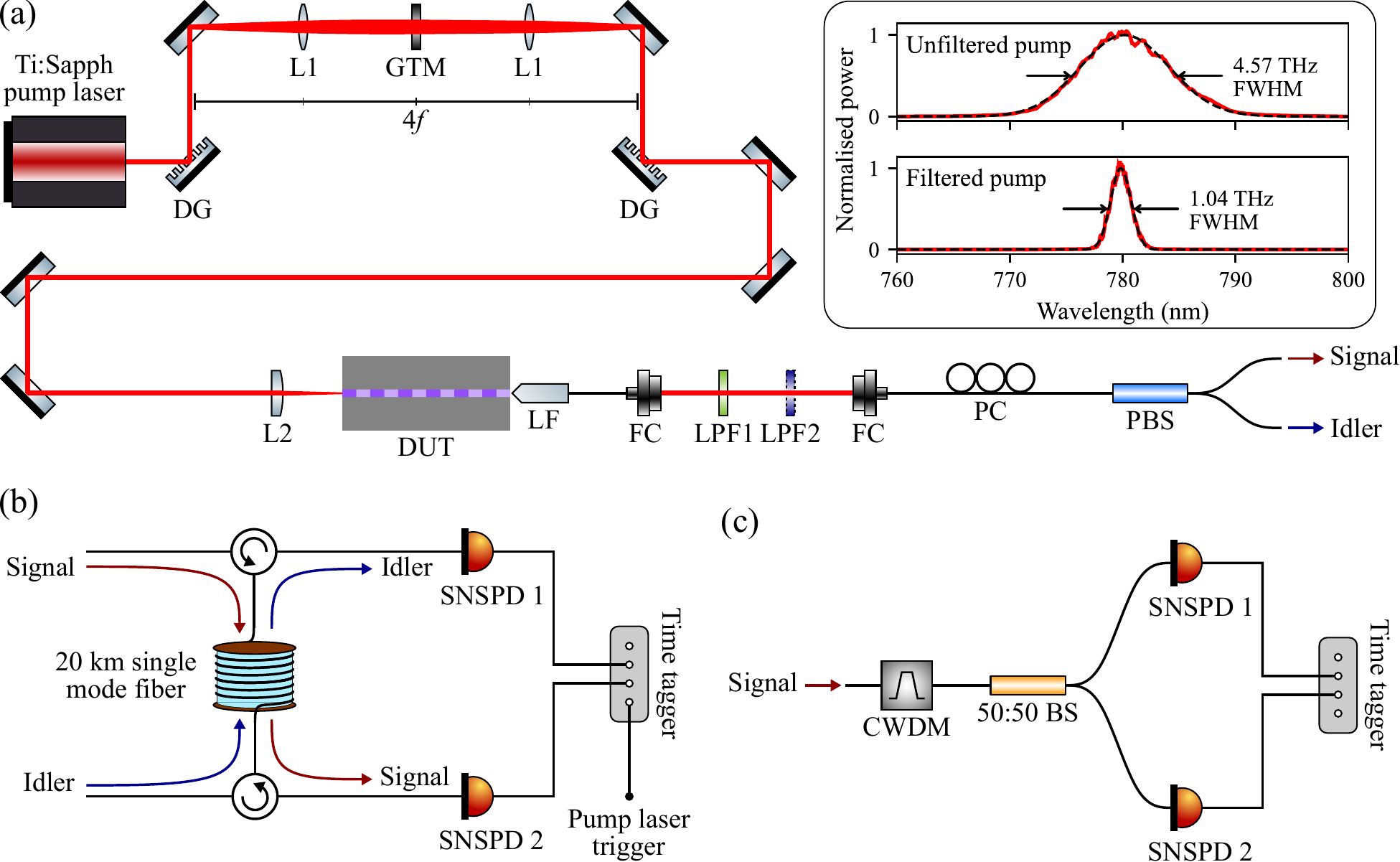}
    \caption{
        Setup for characterizing the biphoton state generated via SPDC. 
        (a)~Pump filtering and waveguide in- and out-coupling setup. 
        The spectrum of a 775~nm-wavelength pulse generated by a Ti:sapphire laser is modified using a pair of diffraction gratings (DG) and a Gaussian transmission filter (GTF) in a $4f$-configuration.
        Insets: normalized pump pulse spectra without (upper panel) and with (lower panel) filtering measured using an optical spectrum analyzer, black dashed lines are Gaussian fits.
        L1,~lens with focal length $f$; 
        GTM,~Gaussian transmission mask; 
        L2,~in-coupling aspheric lens ($4.5$ $\si{\mm}$ focal length, 0.55 numerical aperture); 
        LF,~lensed fiber; 
        FC,~fiber coupler; 
        LPF1,~long-pass filter for pump filtering; 
        LPF2,~long-pass filter for filtering the idler photon used in the unheralded second-order correlation measurement only; 
        PC,~polarization controller; 
        PBS,~fiber polarizing beamsplitter. 
        (b)~Fiber spectroscopy schematic for JSI measurement. 
        Signal and idler photons counterpropagate through single-mode fiber of a $20~\si{\km}$ length using circulators.
        Arrival times of the signal and idler photons at superconducting nanowire single-photon detectors (SNSPDs) are recorded relative to an electrical pulse generated by the pump pulse using time tagging electronics. 
        (c)~Hanbury-Brown-Twiss (HBT) setup for measuring the unheralded second-order correlation function $g^{(2)}$ of the signal beam. 
        An optional coarse-wavelength division multiplexer channel (CWDM) at $\lambda_0 = 1591.08~\si{\nm}$ ($15.6~\si{\nm}$ 3~dB bandwidth) is used to coarsely filter the signal photon.
    }
    \vspace{-4ex}
    \label{fig:setups}
\end{figure}

Next, we characterize the spectral separability of the biphoton state produced directly via SPDC in the aforementioned waveguide with Gaussian-apodized QPM. 
This is done in two ways: dispersive fiber spectroscopy to quantify the JSI of the state and estimate heralded-state spectral purity~\cite{Chen:17}, and an unheralded second-order correlation $g^{(2)}$ measurement of the signal beam to directly estimate the heralded-state purity \cite{Christ:11} (Fig.~\ref{fig:setups}).
Following Ref.~\cite{Chen:19}, a pair of diffraction gratings and a Gaussian transmission mask are used to produce a pump pulse centered on $\lambda_p = 780~\si{\nm}$ with $\Delta \nu = 1.04~\si{\THz}$ (Fig.~\ref{fig:setups}a, inset), which is close to the simulated optimized pump profile (Fig.~\ref{fig:sfg_expt}b,~IV).
The pump beam is coupled into the device using an aspheric lens to avoid the dispersion induced by a lensed fiber. 
The generated signal and idler photons are collected using a lensed fiber and separated using a fiber polarizing beamsplitter following a long-pass filter which filters out the pump beam (Fig.~\ref{fig:setups}a). 
By means of a pair of optical circulators, the JSI of the biphoton state is characterized by counterpropagating the signal and idler photons over a $20$~$\si{\km}$-long single-mode fiber (Fig.~\ref{fig:setups}b).
Due to fiber dispersion, $D_\lambda \sim 17~\si{\ps/\nm/\km}$, the wavelength distribution of the signal and idler photons are mapped to an arrival-time delay of $\Delta t(\lambda) = D_\lambda \cdot L \cdot \qty(\lambda - \lambda_0)$, where $\lambda_0 = 1.55~\si{\micro\m}$. 
Thus, the joint histogram of the arrival time delays of the signal and idler photons, relative to an electrical pulse generated concurrently with the pump pulse, is used to determine the JSI. 
% JSI measurement figure
\begin{figure}[!htb]
    \centering
    \includegraphics[width=0.95\linewidth]{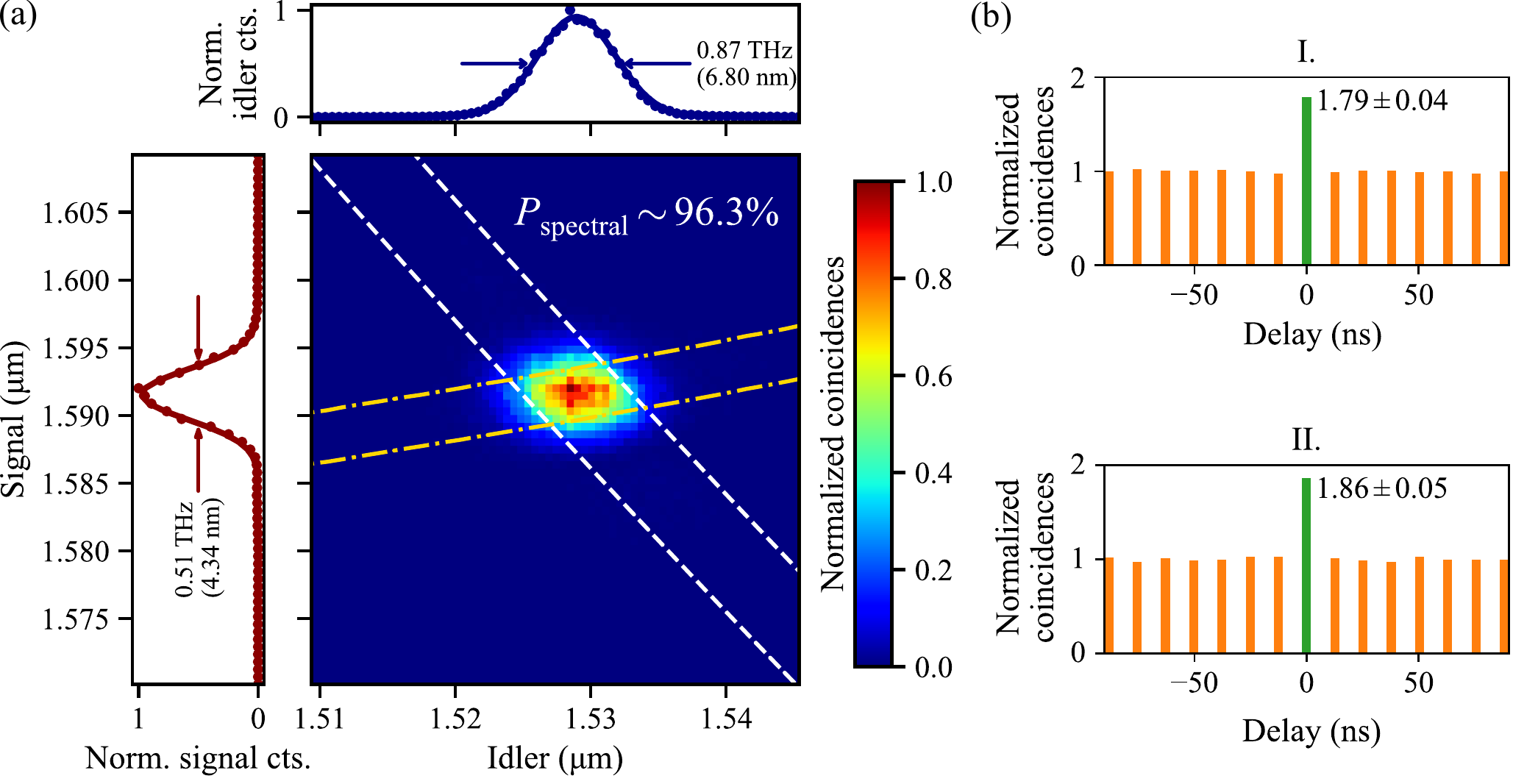}
    \caption{
        Measured JSI and time-integrated $g^{(2)}$ for a biphoton state generated by SPDC. 
        (a)~JSI measured using fiber dispersion-based time-of-flight spectroscopy.  
        Overlaid dashed white (dashed-dot gold) lines indicate the pump envelope (PMF) FWHM extracted from the SFG measurement. 
        Panel above (left of) the JSI is the normalized spectrum of the idler (signal) beam. 
        (b)~Unheralded normalized coincidence histograms for (I)~unfiltered signal beam and (II)~coarsely filtered signal beam.
        Coincidence events per bin integrated over a $4~\si{\ns}$ window and normalized to the mean detection events per bin at non-zero delay.
        The zero-delay bin is highlighted in green.
        Peaks are separated in time according to the repetition rate, $f_r = 79.4~\si{\MHz}$, of the Ti:sapphire laser. 
    }
    \vspace{-3ex}
    \label{fig:jsi_expt}
\end{figure}

The measured JSI is consistent with our simulations~(Fig.~\ref{fig:jsi_expt}a). 
Using singular value decomposition of the JSI data~\cite{Zielnicki:18}, we estimate our device to have a spectral purity of $P_\text{spectral} \sim 96.3\%$ in absence of correlations in the phase component of the JSA.
We note that the idler photon bandwidth is slightly narrower than expected while the signal photon bandwidth is broader than expected. 
This bandwidth difference is because the values of $D_\lambda$ in the dispersive fiber at the center wavelengths of the signal and idler photons deviate from the manufacturer's specifications at $1.55~\si{\um}$.
We further note that the measured spectral purity value is slightly higher than that estimated from the product of the measured PMF and a theoretical optimal pump~(Fig.~\ref{fig:sfg_expt}b, panel~IV). 
This is explained collectively by the bandwidth rescaling and the coarse spectral resolution (${\sim} 0.53~\si{\nm}$) of the measurement compared to the finely sampled PMF measurement.
The former resolution is limited by the amount of applied second order dispersion, the pump laser repetition rate, and the timing resolution of our detectors and readout electronics.

Approximately 50~$\micro\text{W}$ of average pump power is estimated to be in the waveguide for the aforementioned experiment.
At this power, we estimate the pair generation probability to be ${\sim}0.015$ per pulse and an on-chip pair generation rate of ${\sim}1.2~\text{MHz}$.
These biphoton generation rates are also in good agreement with numerical estimates based on the measured SFG efficiency. 
The heralding efficiency was however only ${\sim}2\%$ owing to low out-coupling efficiency and loss in the fiber optics components.
Though we hope that improvements to off-chip coupling schemes in this platform~\cite{He:19}, as well as on-chip implementations of pump filtering, PBS, and detector integration, would help improve heralding efficiencies in the future.

To further characterize the biphoton state separability, and thus the heralded-state purity, $P$, we perform a measurement of the unheralded $g^{(2)}$ of the signal beam. 
Unlike the JSI measurement, this measurement is sensitive to spectral correlations resulting from JSA phase~\cite{Christ:11} and also entanglement in other degrees of freedom, and therefore provides additional information about the separability of the generated biphoton state.
After adding a long-pass filter (LPF2) to the setup to block the idler beam~(Fig.~\ref{fig:setups}a), the signal beam is directed to an optional $15.6$~$\si{\nm}$-wide bandwidth filter (BP), which is used to characterize the biphoton state separability with additional coarse spectral filtering, then into a fiber-based Hanbury-Brown-Twiss setup (Fig.~\ref{fig:setups}c). 
A $4$ $\si{\ns}$ coincidence window is used, which far exceeds the pump pulse duration, which is ${\sim} 400~\si{\fs}$, and thus the normalized coincidence rate at zero time delay yields $g^{(2)}$ (Fig.~\ref{fig:jsi_expt}b).
The measured unheralded $g^{(2)}$ are $1.79 \pm 0.04$ and $1.86 \pm 0.05$, for measurements with and without coarse filtering, respectively. 
The purities are thus $P_\text{I}= (79 \pm 4)\%$ and $P_\text{II} = (86 \pm 5)\%$, with the improvement in the latter case owing to suppression of the phasematching pedestal created by poling imperfections~\cite{Pelc:10}.

We note that the measured overall heralded-state purity is less than the $P_\text{spectral} \sim 96.3\%$ estimated from the JSI measurement. 
This is possibly due to unintended entanglement in polarization and spatial degrees of freedom in our type-II process, and phase chirp in the pump pulse leading to JSA phase correlations.
We believe higher purities will be attainable with further refinement of our fabrication process, though bulk sources with spectral purities comparable to this work have already been used to demonstrate high efficiency entanglement swapping and teleportation~\cite{Jin2015}.
Moreover, the heralded-state purity of our SPDC-based biphoton source is, to our knowledge, the highest of any unfiltered single-pass lithium niobate-based approach.

In conclusion, we have demonstrated record-high spectral separability of a single-pass, lithium niobate-based photon pair source in the telecommunications band, without any narrow spectral filtering, using a thin-film lithium niobate waveguide. 
This work is a valuable addition to this rapidly emerging thin-film platform's unique capabilities and indicates its potential to form the basis of scalable, integrated quantum photonics. 
 
\begin{backmatter}
\bmsection{Funding}
National Science Foundation (OMA-2137723, CCF-1918549, DMR-1231319, ECCS-1839197, ECCS-1542152, ECCS-2026822, EEC-1941583, OIA-2040695, ECCS-1541959, EFMA-1741651); U.S. Department of Energy (DE-SC0020376, DE-AC02-76SF00515); Army Research Office (W911NF2010248); National Center for Research Resources (S10RR02557401); Nippon Telegraph and Telephone (NTT Research 146395); Harvard Quantum Initiative (HQI Seed Funding).

\bmsection{Acknowledgments} 
The authors thank M.~Yeh, S.~Ghosh, M.~Jankowski, M.~Yu, B.~Desiatov, and P.~G.~Kwiat for helpful discussions and assistance with the experiment. N.S. acknowledges funding from the AQT Intelligent Quantum Networks and Technologies (INQNET) research program.

\bmsection{Disclosures}
M.L.: HyperLight Corporation (I,S).

\bmsection{Data availability statement} 
Data underlying the results presented in this paper are not publicly available at this time but may be obtained from the authors upon reasonable request. 

% \bmsection{Supplemental document}
% \hbox{See \href{[insert link here]}{Supplement~1} for supporting content.}
\end{backmatter}

\smallskip
\noindent\textsuperscript{\textdagger}These authors contributed equally to this work.

% Bibliography
\bibliography{bibliography}

\begin{thebibliography}{10}
\newcommand{\enquote}[1]{``#1''}

\bibitem{Xu:22}
M.~Xu, Y.~Zhu, F.~Pittal\`{a}, J.~Tang, M.~He, W.~C. Ng, J.~Wang, Z.~Ruan,
  X.~Tang, M.~Kuschnerov, L.~Liu, S.~Yu, B.~Zheng, and X.~Cai,
  \enquote{Dual-polarization thin-film lithium niobate in-phase quadrature
  modulators for terabit-per-second transmission,}
  {\protect\JournalTitle{Optica}} \textbf{9}, 61--62 (2022).

\bibitem{Wang:18}
C.~Wang, C.~Langrock, A.~Marandi, M.~Jankowski, M.~Zhang, B.~Desiatov, M.~M.
  Fejer, and M.~Lon\v{c}ar, \enquote{Ultrahigh-efficiency wavelength conversion
  in nanophotonic periodically poled lithium niobate waveguides,}
  {\protect\JournalTitle{Optica}} \textbf{5}, 1438--1441 (2018).

\bibitem{Desiatov:19}
B.~Desiatov and M.~Lončar, \enquote{Silicon photodetector for integrated
  lithium niobate photonics,} {\protect\JournalTitle{Applied Physics Letters}}
  \textbf{115}, 121108 (2019).

\bibitem{Elkus:19}
B.~S. Elkus, K.~Abdelsalam, A.~Rao, V.~Velev, S.~Fathpour, P.~Kumar, and G.~S.
  Kanter, \enquote{Generation of broadband correlated photon-pairs in short
  thin-film lithium-niobate waveguides,} {\protect\JournalTitle{Opt. Express}}
  \textbf{27}, 38521--38531 (2019).

\bibitem{Ma:2020}
Z.~Ma, J.-Y. Chen, Z.~Li, C.~Tang, Y.~M. Sua, H.~Fan, and Y.-P. Huang,
  \enquote{Ultrabright quantum photon sources on chip,}
  {\protect\JournalTitle{Phys. Rev. Lett.}} \textbf{125}, 263602 (2020).

\bibitem{Javid:21}
U.~A. Javid, J.~Ling, J.~Staffa, M.~Li, Y.~He, and Q.~Lin,
  \enquote{Ultrabroadband entangled photons on a nanophotonic chip,}
  {\protect\JournalTitle{Phys. Rev. Lett.}} \textbf{127}, 183601 (2021).

\bibitem{Eisaman:11}
M.~D. Eisaman, J.~Fan, A.~Migdall, and S.~V. Polyakov, \enquote{Invited review
  article: Single-photon sources and detectors,} {\protect\JournalTitle{Review
  of Scientific Instruments}} \textbf{82}, 071101 (2011).

\bibitem{Mosley:08}
P.~J. Mosley, J.~S. Lundeen, B.~J. Smith, P.~Wasylczyk, A.~B. U'Ren,
  C.~Silberhorn, and I.~A. Walmsley, \enquote{Heralded generation of ultrafast
  single photons in pure quantum states,} {\protect\JournalTitle{Phys. Rev.
  Lett.}} \textbf{100}, 133601 (2008).

\bibitem{URen:05}
A.~U'Ren, C.~Silberhorn, K.~Banaszek, I.~Walmsley, R.~Erdmann, W.~Grice, and
  M.~Raymer, \enquote{Generation of pure-state single-photon wavepackets by
  conditional preparation based on spontaneous parametiric downconversion,}
  {\protect\JournalTitle{Laser Physics}} \textbf{15}, 146--161 (2005).

\bibitem{MeyerScott:17}
E.~Meyer-Scott, N.~Montaut, J.~Tiedau, L.~Sansoni, H.~Herrmann, T.~J. Bartley,
  and C.~Silberhorn, \enquote{Limits on the heralding efficiencies and spectral
  purities of spectrally filtered single photons from photon-pair sources,}
  {\protect\JournalTitle{Phys. Rev. A}} \textbf{95}, 061803 (2017).

\bibitem{Luo:15}
K.-H. Luo, H.~Herrmann, S.~Krapick, B.~Brecht, R.~Ricken, V.~Quiring, H.~Suche,
  W.~Sohler, and C.~Silberhorn, \enquote{Direct generation of genuine
  single-longitudinal-mode narrowband photon pairs,} {\protect\JournalTitle{New
  Journal of Physics}} \textbf{17}, 73039 (2015).

\bibitem{Guo2016}
X.~Guo, C.~ling Zou, C.~Schuck, H.~Jung, R.~Cheng, and H.~X. Tang,
  \enquote{Parametric down-conversion photon-pair source on a nanophotonic
  chip,} {\protect\JournalTitle{Light: Science {\&} Applications}} \textbf{6},
  e16249--e16249 (2016).

\bibitem{Graffitti2018}
F.~Graffitti, P.~Barrow, M.~Proietti, D.~Kundys, and A.~Fedrizzi,
  \enquote{Independent high-purity photons created in domain-engineered
  crystals,} {\protect\JournalTitle{Optica}} \textbf{5}, 514 (2018).

\bibitem{Chen:19}
C.~Chen, J.~E. Heyes, K.-H. Hong, M.~Y. Niu, A.~E. Lita, T.~Gerrits, S.~W. Nam,
  J.~H. Shapiro, and F.~N.~C. Wong, \enquote{Indistinguishable single-mode
  photons from spectrally engineered biphotons,} {\protect\JournalTitle{Opt.
  Express}} \textbf{27}, 11626--11634 (2019).

\bibitem{Jankowski2021}
M.~Jankowski, J.~Mishra, and M.~M. Fejer, \enquote{Dispersion-engineered
  $\chi^{(2)}$ nanophotonics: a flexible tool for nonclassical light,}
  {\protect\JournalTitle{Journal of Physics: Photonics}} \textbf{3}, 042005
  (2021).

\bibitem{Huang:06}
J.~Huang, X.~P. Xie, C.~Langrock, R.~V. Roussev, D.~S. Hum, and M.~M. Fejer,
  \enquote{Amplitude modulation and apodization of quasi-phase-matched
  interactions,} {\protect\JournalTitle{Opt. Lett.}} \textbf{31}, 604--606
  (2006).

\bibitem{Chen:17}
C.~Chen, C.~Bo, M.~Y. Niu, F.~Xu, Z.~Zhang, J.~H. Shapiro, and F.~N.~C. Wong,
  \enquote{Efficient generation and characterization of spectrally factorable
  biphotons,} {\protect\JournalTitle{Opt. Express}} \textbf{25}, 7300--7312
  (2017).

\bibitem{Christ:11}
A.~Christ, K.~Laiho, A.~Eckstein, K.~N. Cassemiro, and C.~Silberhorn,
  \enquote{Probing multimode squeezing with correlation functions,}
  {\protect\JournalTitle{New Journal of Physics}} \textbf{13}, 033027 (2011).

\bibitem{Laudenbach:16}
F.~Laudenbach, H.~H\"{u}bel, M.~Hentschel, P.~Walther, and A.~Poppe,
  \enquote{Modelling parametric down-conversion yielding spectrally pure photon
  pairs,} {\protect\JournalTitle{Opt. Express}} \textbf{24}, 2712--2727 (2016).

\bibitem{Zielnicki:18}
K.~Zielnicki, K.~Garay-Palmett, D.~Cruz-Delgado, H.~Cruz-Ramirez, M.~F.
  O'Boyle, B.~Fang, V.~O. Lorenz, A.~B. U'Ren, and P.~G. Kwiat, \enquote{Joint
  spectral characterization of photon-pair sources,}
  {\protect\JournalTitle{Journal of Modern Optics}} \textbf{65}, 1141--1160
  (2018).

\bibitem{He:19}
L.~He, M.~Zhang, A.~Shams-Ansari, R.~Zhu, C.~Wang, and M.~Lon\v{c}ar,
  \enquote{Low-loss fiber-to-chip interface for lithium niobate photonic
  integrated circuits,} {\protect\JournalTitle{Optics Letters}} \textbf{44},
  2314--2317 (2019).

\bibitem{Pelc:10}
J.~S. Pelc, C.~Langrock, Q.~Zhang, and M.~M. Fejer, \enquote{Influence of
  domain disorder on parametric noise in quasi-phase-matched quantum frequency
  converters,} {\protect\JournalTitle{Opt. Lett.}} \textbf{35}, 2804--2806
  (2010).

\bibitem{Jin2015}
R.-B. Jin, M.~Takeoka, U.~Takagi, R.~Shimizu, and M.~Sasaki, \enquote{Highly
  efficient entanglement swapping and teleportation at telecom wavelength,}
  {\protect\JournalTitle{Scientific Reports}} \textbf{5} (2015).

\end{thebibliography}
 % Full bibliography added automatically for Optics Letters submissions; the following line will simply be ignored if submitting to other journals.
% Note that this extra page will not count against page length
\bibliographyfullrefs{bibliography}
\end{document}